\documentclass{appolb}
\usepackage{graphicx}

\begin{document}
\title{Proton Spin in Deep Inelastic Scattering
\thanks{Presented at the workshop on Diffraction and Low-x, Reggio Calabria, Aug. 26 - Sept. 1, 2018}%
}
\author{B.~Povh
\address{ Max-Planck Institut f\"ur Kernphysik, 69029 Heidelberg, Germany}\vspace{0.5cm}
\\
{M.~Rosina
}
\address{Faculty of Math and Physics, University of Ljubljana\\
J.~Stefan Institute, Ljubljana, Slovenia}\vspace{0.5cm}
}
\maketitle
\begin{abstract}
So far the analyses of the polarized structure functions of the proton and 
neutron have been limited to
the evaluation of their integrals and comparing them to the prediction of the
static quark model of the nucleon. 
We extended our analysis
to the x dependence of the polarized structure functions and observe:
the measured structure function excellently agrees with the prediction of the
static quark model for
Bjorken $x>0.2$
and drops rapidly for $x<0.2$. It is suggested that for Bjorken $x>0.2$ electrons
get scattered on the undamaged constituent quarks (alias valence quarks) denoted as quasi-elastic scattering
on the constituent quarks and for $x<0.2$ the
constituent quarks fragment. In the fragmentation strong interaction is involved which
does not preserve the polarization.
\end{abstract}
\section{Introduction}
The weak decays of the baryon octet are well reproduced in the flavor SU3.
The weak vector current transition is given by the Fermi coupling constant $G_F$ multilied by
the cosine of the Cabbibo angle $\cos \theta_C$ for the neutron decay and by $\sin \theta_C$
for the hyperons decays. For the axial-vector transition (partially conserving axial-vactor current)
two experimental coupling constants $g_A$ for the neutron decay and $g_{\Sigma}$ for the
hyperons decays have to be used in order to restore the SU3 symmetry \cite{Cabbibo}
for this decay. The two coupling
constants for constituent quarks  are smaller than the coupling constants for the elementary quarks, witnessing
that the angular momentum of the constituent quark is not carried entirely by the quark spin.
Knowing the axial-vector transitions of the hyperons and neutron the spin carried by quarks
in the baryon octet is uniquely determined. By means of the light cone algebra,
taking into account the moowing system and 
assuming that the strange quarks do not contribute to the polarization of the proton and neutron
Ellis and Jaffe \cite{Ellis} calculated the integrals of the polarized structure function for the proton. 
\begin {equation}
\int g^p_1(x)dx=\frac{g_A}{6}(1-b)= 0.175
\label{one}
\end{equation}
and for the neutron:
\begin{equation}
\int g^n_1(x)dx=-\frac{g_A}{6} b= 0.023.
\label{two}
\end{equation}
In the two equations the parameter $b$ reduces the integrals as the concequence  the fluctuation p$\rightarrow$n+$\pi^+$
and n$\rightarrow$p+$\pi^-$. 
Since the first measurement by the EMC collaboration in 1989 \cite{EMC} and following experiments
 of the SMC collaboration \cite{SMC}, the NMC collaboration \cite{NMC}, the HERMES collaboration \cite{HERMES} and the COMPAS collaboration \cite{COMPAS} the integral of the proton polarized structure function strongly
disagrees with the predicted value. The HERMES value for the integral is: 
\begin{equation}
\int g^p_1(x)dx=0.127\pm0.002 \pm0.007\pm0.005.
\end{equation}
\section{Dependence of the polarized structure function on the Bjorken x}
The difference between the quark polarization calculated on the light cone and in the rest frame of the 
nucleon is marginal. So we sketch the derivation of the quark polarization in the rest frame
of the nucleon. For the three elementary uud quarks the integral of the quark polarization is
\begin{equation}
<p\uparrow|\Sigma\sigma_{zi}|p\uparrow>=2\frac{g_A}{6}<p|p>
\label{four}
\end{equation}
and $g_A=\frac{5}{3}$ as can be found in any text book of particle physics, for instant \cite{Close}.
Identical expression (\ref{four}) is valid for the restored SU3 if for $g_A=1.27$ is taken and the measured wave function for the valence quarks is used. 
The proportionality between the polarized and not polarized structure functions
is valid in all models in which the 3-quark wave function  factorizes in 
color $\times$ orbital $\times$ spin-isospin parts.
The left side of (\ref{four}) corresponds to 
twice the integral over the polarized structure function, the right one to the integral over 
the structure function of the valence quarks multiplied with the reducing factor. 
Omitting the integrals the polarized structure function sounds:
\begin{equation}
xg^p_2(x)=\frac{g_A}{6} F^{p(val)}_2(x)
\end{equation}
There is no direct measurement of the valence-quark structure function. With a single
measurement of the structure function to single out the 
valence-quarks structure function is not possible. For the fit too many
parameters have to be assumed ad hoc. Particularly the assupmtion
for the ratio 2:1 for the u and d valence quarks used in all the fits neglects
the pion fluctuation and leads to unrealistic results.  
The best
reconstruction of the valence-quarks structure function taking into account the pion fluctuation can be obtained from the measurement of the Gottfried sum rule (\ref{six})             
\begin{equation}
\int\frac{1}{x}(F^p_2(x)-F^n_2(x))=\frac{1}{3}(1-2a)
\label{six}
\end{equation}
and $a$ is the probability for the $p\rightarrow n+\pi^+$ fluctuation.
\begin{figure}[h]
\centerline{%
\includegraphics[width=7.0cm]{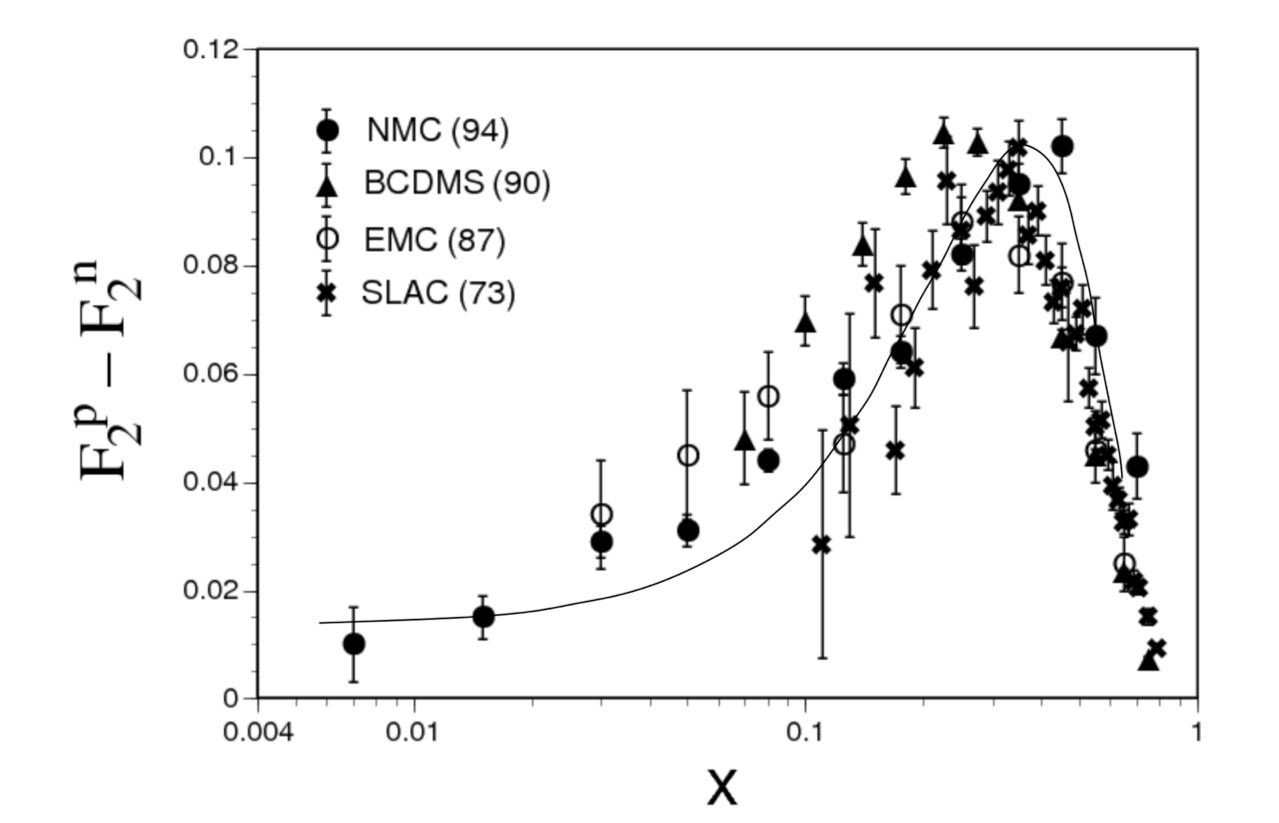}}
\caption{Difference between the proton and neutron structure function.
The fit has been done for the NMC data.}
\label{Fig:fig1}
\end{figure}
In (Fig.\ref{Fig:fig1}) we show the fit to the NMC data. From the equation (\ref{six}) we see that
the pion fluctuation is deduced twice, once taken off the proton and ones shifted to the neutron.
From the equations (\ref{one} and \ref{two}) one sees how the missing integrals
over the proton structure function appears in the integral of the neutron. 
The missing part of the structure function due to the pion fluctuation can be 
credible restore by the measured neutron structure function. The full 
\begin{figure}[t]
\centerline{%
\includegraphics[width=7.0cm]{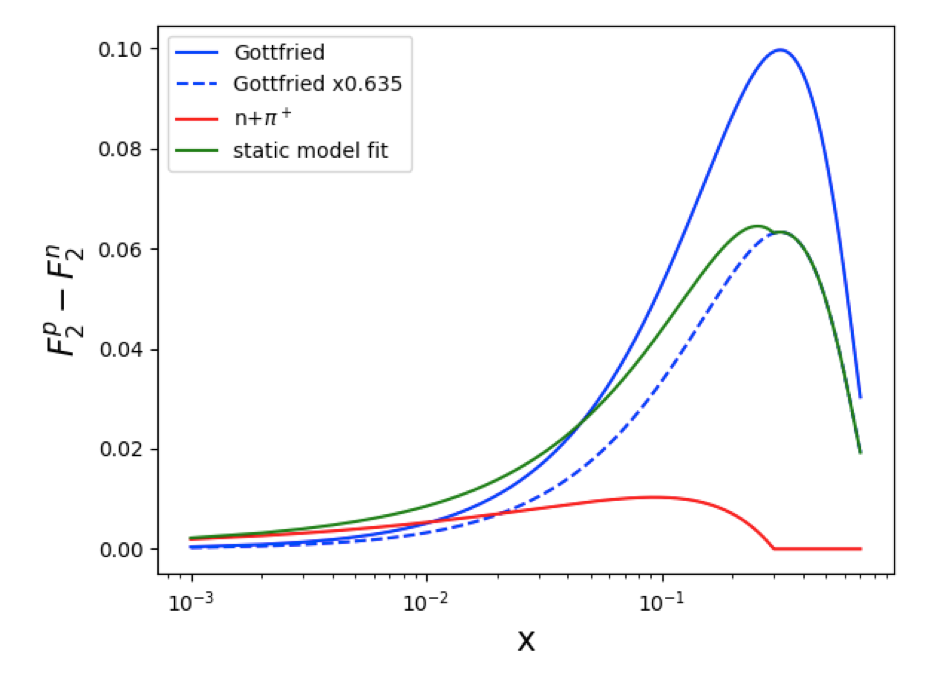}}
\caption{Reconstructure  of the polarized structure function of the static model.}
\label{Fig:fig2}
\end{figure}
reconstruction is shown in Fig.\ref{Fig:fig2}. 
The following Fig.\ref{Fig:fig3} shows the comparison between the
prediction of the static model \cite{Ellis} and the polarized structure 
function by the experimental data
\begin{figure}[h]
\centerline{%
\includegraphics[width=7.0cm]{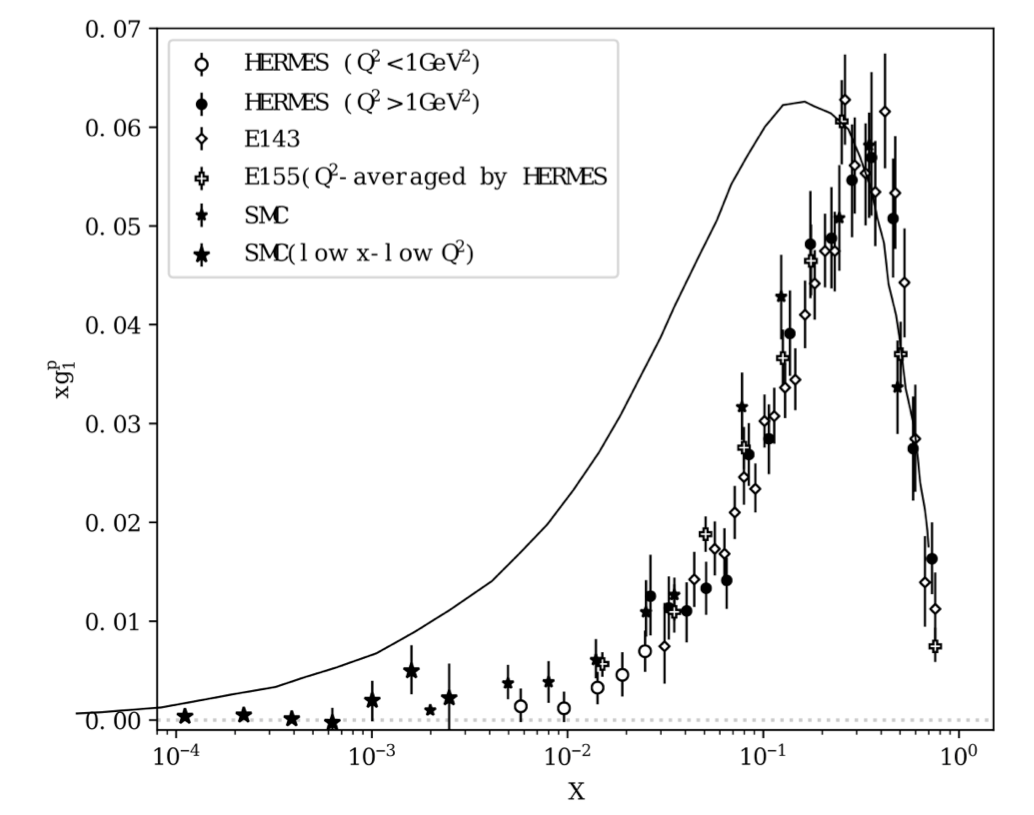}}
\caption{Comparison of the prediction of the statical model and the data.}
\label{Fig:fig3}
\end{figure}
In fact the full reconstruction of the predicted polarized structure 
was not necessary as we only wonted to show that for x$>$ 0.2 the polarized
structure function obtains the maximum possible value. On other hand 
it is good as the full reconstruction
shows that the the integral over the polarized structure function amounts
for about half of the predicted integral in agreement with data.
\section{Discussion and Conclusion}
\begin{figure}[h]
\centerline{%
\includegraphics[width=9.0cm, angle=-0.5]{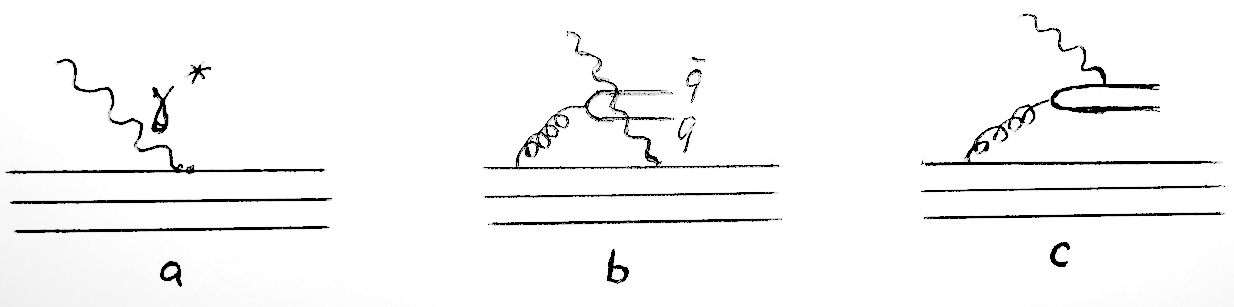}}
\caption{(a) quasi-elastic scattering on the constituent quark, (b) scattering
on a fraction of the constituent quark and (c) scattering on the quark-antiquark
pairs.}
\label{Fig:fig4}
\end{figure}

 Fig.(\ref{Fig:fig3}) is strongly suggestive. At Bjorken $x=\frac{1}{3}$ one
expects that the electrons get elastically scattered on the objects with a mass of one third 
of the nucleon mass. It is the  rather plausible to identify the events
which conserve the quark spin with the elastic scattering of electrons on the 
bound constituent quarks denoted usually as quasi elastic scattering.
In Fig.\ref{Fig:fig4} the quasi elastic scattering on an undamaged constituent quark, 
scattering on a fraction of a constituent quark and scattering  on the quark-antiquark pairs
is shown.
In the interpretation of the DIS measurements it is assumed that the nucleon
is completely dissolved in the current quarks and gluonen. From the identification
of the valence quarks in the structure functions it is, however, obvious that they are composed
objects and only as the wholes carry the spin of the constituent quark. The undamaged 
constituent quark carries the spin of the current quark the damaged
not necessarily. Even if the current quark of the damaged quark
conserves the polarization the damaged constituent quark as a whole it does not.
There is further an important information that follows  from the analysis of the polarization
measurements in DIS.
In the scattering experiments the two scales of the hadrons are clearly demonstrated
by the interplay between the soft and hard interaction as summarized in \cite{boris1}
and reported earlier in \cite{boris2} and \cite{boris3}. The interaction 
involving hadron substructure which is
responsible for the hard interaction is dominated by the 
gluon exchange.  Therefore the members of the substructure were called gluon spots
\cite{boris2}. It is rather 
obvious to identify the gluon spots with the constituent quarks.
Identifying the substructure of the light hadrons with the constituent quarks 
gives the models with three constituent quarks in a common mean field
theoretical justification.\\
{Acknowledgments:} We are thankful to Boris Kopeliovich for the
critical discussions of the paper and pointing us out the close connection of the
spin measurements in DIS and the substructure of the hadrons in 
scattering experiments. Thomas Walcher participated in this study
at the early stage and made very able contribution.


\begin{thebibliography}{99}
\bibitem{Cabbibo} 
N.~Cabbibo, E.~C.~Swallow and R.~Winston, {\it Ann. Rev. Nucl. Part. Sci.} {\bf 53} 39 (2003).
\bibitem{Ellis}
J.~Ellis and R.~Jaffe, {\it Phys. Rev.} {\bf D9} 1444 (1974)
\bibitem{EMC} 
J.~Ashman et al. {\it Nucl. Phys.}  {\bf B328} 1 (1989)
\bibitem{SMC}
SMC collaboration, D.~Adeva et al. {\it Phys. Rev.} {\bf D58} 112001 (1998)
\bibitem{NMC}
NMC collaboration, P.~Amaudruz et al. {\it Phys. Rev. Lett. } {\bf 66} 21 (1991)
\bibitem{HERMES}
HERMES collaboration, A.~Airapetian et al. {\it  Phys. Rev. } {\bf D75} 012007 (2007)
\bibitem{COMPAS}
COMPASS collaboration, C.~Adolph et al.{\it Phys. Lett.}{\bf B753} 18 (2016) 1503.08935   
\bibitem{Close}
F.~E.~Close {Quarks and Patons}
\bibitem{boris1}
B.~Kopeliovich, I.~Potashnikova, B.~Povh and I.~Schmitt {\it Phys. Rev.} {\bf D76}
094020 (2007)
\bibitem{boris2}
B.~Kopeliovich and B.~Povh {\it J. Phys.}{\bf G30} 5999 (2004)
\bibitem{boris3}
B.~Kopeliovich, A.~Sch\"afer, A.~Tarasov {\it Phys. Rev.}{\bf D62} 054022 (2000)
\end{thebibliography}
\end{document}